\documentclass[runningheads]{llncs}
\usepackage[T1]{fontenc}
\usepackage{graphicx}
\usepackage{amsmath}
\usepackage{multirow}
\usepackage{booktabs}
\usepackage{makecell}
\usepackage{array} 
\usepackage{xcolor}

\usepackage{pifont}
\usepackage{amssymb}
\usepackage{hyperref}
\hypersetup{hypertex=true,
colorlinks=true,
linkcolor=blue,
anchorcolor=blue,
citecolor=blue,
}

\begin{document}
\title{Dual Arbitrary Scale Super-Resolution for Multi-Contrast MRI}
\author{Jiamiao Zhang\inst{1} \and
Yichen Chi\inst{1} \and
Jun Lyu\inst{2} \and
Wenming Yang\inst{1} \and
Yapeng Tian\inst{3}}

\authorrunning{J. Zhang et al.}

\institute{Shenzhen International Graduate School, Tsinghua University,
Beijing, China\\
\email{yang.wenming@sz.tsinghua.edu.cn}\\
\and
School of Nursing, The Hong Kong Polytechnic University, Hong Kong\\
\and
Department of Computer Science, The University of Texas at Dallas, Richardson, USA\\
\url{https://github.com/jmzhang79/Dual-ArbNet} 
}
\maketitle              
\begin{abstract}
Limited by imaging systems, the reconstruction of Magnetic Resonance Imaging (MRI) images from partial measurement is essential to medical imaging research. Benefiting from the diverse and complementary information of multi-contrast MR images in different imaging modalities, multi-contrast Super-Resolution (SR) reconstruction is promising to yield SR images with higher quality. In the medical scenario, to fully visualize the lesion, radiologists are accustomed to zooming the MR images at arbitrary scales rather than using a fixed scale, as used by most MRI SR methods. In addition, existing multi-contrast MRI SR methods often require a fixed resolution for the reference image, which makes acquiring reference images difficult and imposes limitations on arbitrary scale SR tasks. To address these issues, we proposed an implicit neural representations based dual-arbitrary multi-contrast MRI super-resolution method, called Dual-ArbNet. First, we decouple the resolution of the target and reference images by a feature encoder, enabling the network to input target and reference images at arbitrary scales. Then, an implicit fusion decoder fuses the multi-contrast features and uses an Implicit Decoding Function~(IDF) to obtain the final MRI SR results. Furthermore, we introduce a curriculum learning strategy to train our network, which improves the generalization and performance of our Dual-ArbNet. Extensive experiments in two public MRI datasets demonstrate that our method outperforms state-of-the-art approaches under different scale factors and has great potential in clinical practice. 


\end{abstract}
\section{Introduction}
Magnetic Resonance Imaging (MRI) is one of the most widely used medical imaging modalities, as it is non-invasive and capable of providing superior soft tissue contrast without causing ionizing radiation. However, it is challenging to acquire high-resolution MR images in practical applications~\cite{feng2021brain} due to the inherent shortcomings of the systems~\cite{plenge2012super,van2012super} and the inevitable motion artifacts of the subjects during long acquisition sessions.

Super-resolution (SR) techniques are a promising way to improve the quality of MR images without upgrading hardware facilities. Clinically, multi-contrast MR images, {\itshape e.g.}, T1, T2 and PD weighted images are obtained from different pulse sequences~\cite{liu2021regularization,sun2020extracting}, which can provide complementary information to each other~\cite{chen2015accuracy,feng2021multi}. Although weighted images reflect the same anatomy, they excel at demonstrating different physiological and pathological features. Different time is required to acquire images with different contrast. In this regard, it is promising to leverage an HR reference image with a shorter acquisition time to reconstruct the modality with a longer scanning time. Recently, some efforts have been dedicated to multi-contrast MRI SR reconstruction. Zeng {\itshape et al.} proposed a deep convolution neural network to perform single- and multi-contrast SR reconstruction~\cite{zeng2018simultaneous}. Dar {\itshape et al.} concatenated information from two modalities into the generator of a generative adversarial network(GAN)~\cite{dar2020prior}, and Lyu {\itshape et al.} introduced a GAN-based progressive network to reconstruct multi-contrast MR images~\cite{lyu2020multi}. Feng {\itshape et al.} used a multi-stage feature fusion mechanism for multi-contrast SR~\cite{feng2021multi}. Li {\itshape et al.} adopted a multi-scale context matching and aggregation scheme to gradually and interactively aggregate multi-scale matched features~\cite{li2022transformer}. 
Despite their effectiveness, these networks impose severe restrictions on the resolution of the reference image, largely limiting their applications. In addition, most existing multi-contrast SR methods only work with fixed integer scale factors and treat different scale factors as independent tasks. For example, they train a single model for a certain integer scale factor ($\times$2, $\times$4). In consequence, using these fixed models for arbitrary scale SR is inadequate. Furthermore, in practical medical applications, it is common for radiologists to zoom in on MR images at will to see localized details of the lesion. Thus, there is an urgent need for an efficient and novel method to achieve super-resolution of arbitrary scale factors in a single model.

In recent years, several methods have been explored for arbitrary scale super-resolution tasks on natural images, such as Meta-SR~\cite{hu2019meta} and Arb-SR~\cite{wang2021learning}. Although they can perform arbitrary up-sampling within the training scales, their generalization ability is limited when exceeding the training distribution, especially for large scale factors. Inspired by the success of implicit neural representation in modeling 3D shapes~\cite{park2019deepsdf,mescheder2019occupancy,chen2019learning,jiang2020local,sitzmann2020implicit}, several works perform implicit neural representations to the 2D image SR problem~\cite{chen2021learning,nguyen2023single}. Since these methods can sample pixels at any position in the spatial domain, they can still perform well beyond the distribution of the training scale. Also, there is an MRI SR method that combines the meta-upscale module with GAN and performs arbitrary scale SR~\cite{tan2020arbitrary}. However, the GAN-based method generates unrealistic textures, which affects the diagnosis accuracy. 

To address these issues, we propose an arbitrary-scale multi-contrast MRI SR framework. Specifically, we introduce the implicit neural representation to multi-contrast MRI SR and extend the concept of arbitrary scale SR to the reference image domain. Our contributions are summarized as follows:

1) We propose a new paradigm for multi-contrast MRI SR with the implicit neural representation, called Dual-ArbNet. It allows arbitrary scale SR at any resolution of reference images. 

2) We introduce a curriculum learning~\cite{bengio2009curriculum} strategy called Cur-Random to improve the stability, generalization, and multi-contrast fusion performance of the network. 

3) Our extensive experiments can demonstrate the effectiveness of our method. Our Dual-ArbNet outperforms several state-of-the-art approaches on two benchmark datasets: fastMRI~\cite{zbontar2018fastmri} and IXI~\cite{ixi2015dataset}.

\section{Methodology}

\subsection{Background: Implicit Neural Representations}\label{sec:INRBG}
As we know, computers use 2D pixel arrays to store and display images discretely. In contrast to the traditional discrete representation, the Implicit Neural Representation~(INR) can represent an image $I\in R^{H\times W}$ in the latent space $F\in R^{H\times W\times C}$, and use a local neural network ({\itshape e.g.}, convolution with kernel 1) to continuously represent the pixel value at each location. This local neural network fits the implicit function of the continuous image, called Implicit Decoding Function~(IDF). In addition, each latent feature represents a local piece of continuous image~\cite{chen2021learning}, which can be used to decode the signal closest to itself through IDF. Thus, by an IDF $f(\cdot)$ and latent feature $F$, we can arbitrarily query pixel value at any location, and restore images of arbitrary resolution.

\subsection{Network Architecture}
\begin{figure}[tbp]
\centering
\includegraphics[width=1\textwidth]{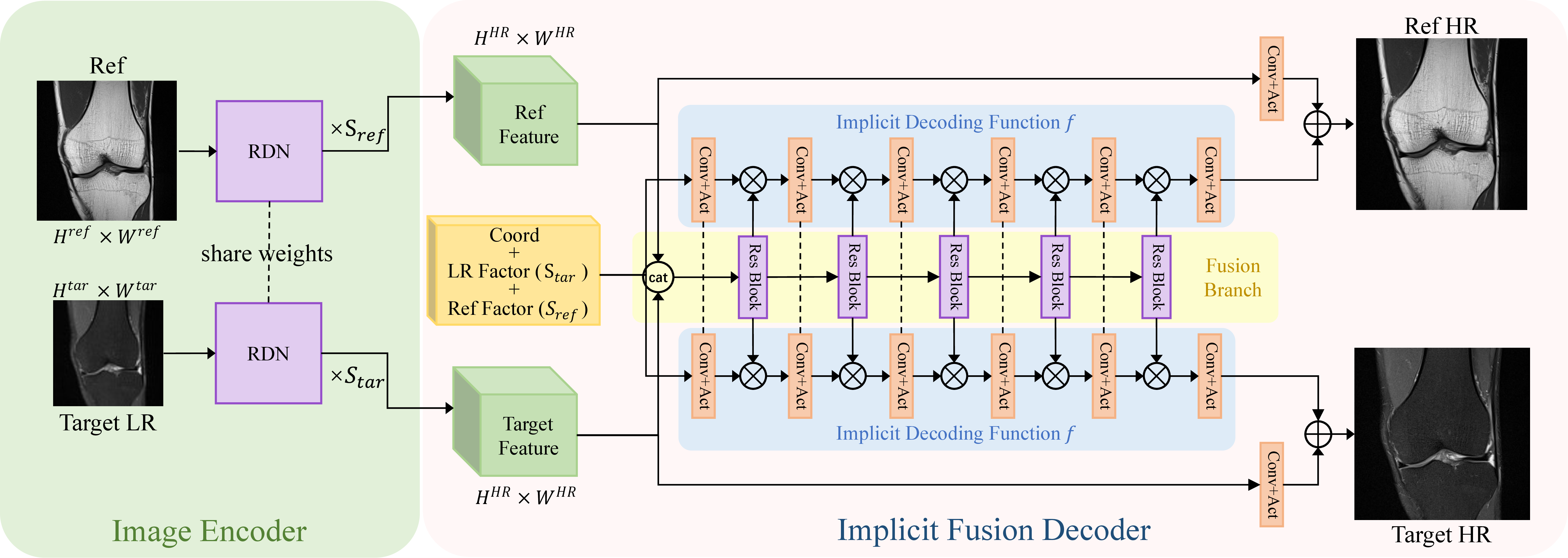}
\caption{Overall architecture of the proposed Dual-ArbNet. Our Dual-ArbNet includes a share-weighted image encoder and an implicit fusion decoder which contains a lightweight fusion branch and an implicit decoding function.
} 
\label{img:arch}
\end{figure}
The overall architecture of the proposed Dual-ArbNet is shown in Fig.\ref{img:arch}. The network consists of an encoder and an implicit fusion decoder. The encoder performs feature extraction and alignment of the target LR and the reference image. The implicit fusion decoder predicts the pixel values at any coordinate by fusing the features and decoding through IDF, thus achieving reconstruction. 

\noindent
\textbf{Encoder.} In the image encoder, Residual Dense Network (RDN)~\cite{zhang2018residual} is used to extract image latent features for the network, and the reference image branch shares weights with the target LR image branch to achieve consistent feature extraction and reduce parameters. To aggregate the neighboring information in the reconstruction process, we further unfold the features of 3×3 neighborhoods around each pixel, expanding the feature channels nine times. 

Since the resolution of target LR and reference image are different, we have to align them to target HR scale for further fusion. With the target image shaped $H^{tar}\times W^{tar}$ and reference image shaped $H^{ref}\times W^{ref}$, we use nearest interpolation to efficiently up-sample their feature maps to the target HR scale $H^{HR}\times W^{HR}$ by two different factors $S_{ref}$ and $S_{tar}$: 
\begin{equation}
\begin{aligned}
   F_{z\uparrow} &= Upsample(RDN(I_{z}), S_{z})
\end{aligned}
\end{equation}
where $z\in \{ref,tar\}$ indicates the reference and target image, $I_{tar}$ and $I_{ref}$ are the input target LR and reference image. In this way, we obtain the latent feature nearest to each HR pixel for further decoding, and our method can handle Arbitrary scale SR for target images with Arbitrary resolution of reference images~(Dual-Arb).

\noindent
\textbf{Decoder.} 
As described in Sec.~\ref{sec:INRBG}, the INR use a local neural network to fit the continuous image representation, and the fitting can be referred to as Implicit Decoding Function~(IDF). In addition, we propose a fusion branch to efficiently fuse the target and reference latent features for IDF decoding. The overall decoder includes a fusion branch and a shared IDF, as shown in Figure~\ref{img:arch}(see right).

Inspired by~\cite{zhang2018residual,woo2018cbam}, to better fuse the reference and target features in different dimensions, we use ResBlock with Channel Attention~(CA) and Spatial Attention~(SA) in our fusion branch. This 5 layers lightweight architecture can capture channel-wise and spatial-wise attention information and fuse them efficiently. The fusion process can be expressed as:
\begin{equation}
\begin{aligned}
   F_{fusion}^{(0)} &= cat(F_{tar\uparrow},F_{ref\uparrow})\\
   F_{fusion}^{(i)}&=L_i(F_{fusion}^{(i-1)})+F_{fusion}^{(i-1)}, & i=1,2,...,5
\end{aligned}
\end{equation}
where $L_i$ indicates the $i$-th fusion layer. 
Then, we equally divide the fused feature $F_{fusion}^{(i)}$ by channel into $F_{fusion,tar}^{(i)}$ and $F_{fusion,ref}^{(i)}$ for decoding respectively.

The IDF in our method is stacked by convolution layer with kernel size 1~($conv_1$) and sin activation function $sin(\cdot)$.
The $conv_1$ and $sin(\cdot)$ are used to transform these inputs to higher dimension space~\cite{nguyen2023single}, thus achieving a better representation of the IDF. Since $conv_1(x)$ can be written as $W\cdot x+b$ without using any adjacent features, this decoding function can query SR value at any given coordinate. 
Akin to many previous works~\cite{chen2021learning,nguyen2023single}, relative coordinate information $P(x,y)$ and scale factors $S_{ref},S_{tar}$ are necessary for the IDF to decode results continuously. At each target pixel $(x,y)$, we only use local fused feature $F_{fusion}$, which represents a local piece of continuous information, and coordinate $P(x,y)$ relative to the nearest fused feature, as well as scale factors $\{S_{ref},S_{tar}\}$, to query in the IDF.
Corresponding to the fusion layer, we stack 6 convolution with activation layers. $i$-th layer's decoding function $f^{(i)}$ can be express as:
\begin{equation}
\begin{aligned}
   f^{(0)}(x,y,z) &= sin\left(W^{(0)}\cdot cat\left(S_{tar},S_{ref},P(x,y)\right)+b^{(0)}\right)\\
   f^{(i)}(x,y,z) &= sin\left(\left(W^{(i)}\cdot f^{(i-1)}(x,y,z)+b^{(i)}\right)\odot F_{fusion,z}^{(i)}(x,y)\right)
\end{aligned}
\end{equation}
where $(x,y)$ is the coordinate of each pixel, and $z\in \{ref,tar\}$ indicates the reference and target image. $\odot$ denotes element-wise multiplication, and $cat$ is the concatenate operation. $W^{(i)}$ and $b^{(i)}$ are weight and bias of $i$-th convolution layer.
Moreover, we use the last layer's output $f^{(5)}(\cdot)$ as the overall decoding function $f(\cdot)$.
By introducing the IDF above, the pixel value at any coordinates $I_{z,SR}(x,y)$ can be reconstructed:
\begin{equation}
\begin{aligned}
   I_{z,SR}(x,y)&=f(x,y,z)+Skip(F_{z\uparrow})
\end{aligned}
\label{eq:imp}
\end{equation}
where $Skip(\cdot)$ is skip connection branch with $conv_1$ and $sin(\cdot)$, $z\in \{ref,tar\}$ . 

\noindent
\textbf{Loss Function.} An L1 loss between target SR results $I_{target,SR}$ and HR images $I_{HR}$ is utilized as reconstruction loss to improve the overall detail of SR images, named as $L_{rec}$.
The reconstructed SR images may lose some frequency information in the original HR images. K-Loss~\cite{zhou2020dudornet} is further introduced to alleviate the problem. Specifically, $K_{SR}$ and $K_{HR}$ denote the fast Fourier transform of $I_{target,SR}$ and $I_{HR}$. In {\itshape k}-space, the value of mask $M$ is set to 0 in the high-frequency cut-off region mentioned in Sec.~\ref{sec:exp}, otherwise set to 1. L2 loss is used to measure the error between $K_{SR}$ and $K_{HR}$. K-Loss can be expressed as:
\begin{equation}
L_{K}=\left\| (K_{SR}-K_{HR})\cdot M \right\|_2
\end{equation}
To this end, the full objective of the Dual-ArbNet is defined as:
\begin{equation}
\begin{aligned}
   L_{full}&= L_{rec}+\lambda _{K}L_{K}
\end{aligned}
\end{equation}
We set $\lambda_{K}=0.05$ empirically to balance the two losses.

\subsection{Curriculum Learning Strategy}
Curriculum learning~\cite{bengio2009curriculum} has shown powerful capabilities in improving model generalization and convergence speed. It mimics the human learning process by allowing the model to start with easy samples and gradually progress to complex samples. 
To achieve this and stabilize the training process with different references, we introduce curriculum learning to train our model, named Cur-Random. This training strategy is divided into three phases, including warm-up, pre-learning, and full-training. Although our image encoder can be fed with reference images of arbitrary resolution, it is more common to use LR-ref~(scale as target LR) or HR-ref~(scale as target HR) in practice. Therefore, these two scales of reference images are used as our settings.

In the warm-up stage, we fix the integer SR scale to integer~(2$\times$, 3$\times$ and 4$\times$) and use HR-Ref to stable the training process. 
Then, in the pre-learning stage, we use arbitrary scale target images and HR reference images to quickly improve the network's migration ability by learning texture-rich HR images. 
Finally, in the full-training stage, we train the model with a random scale for reference and target images, which further improves the generalization ability of the network. 

\section{Experiments}\label{sec:exp}

\noindent
\textbf{Datasets.} Two public datasets are utilized to evaluate the proposed Dual-ArbNet network, including fastMRI~\cite{zbontar2018fastmri}~(PD as reference and FS-PD as target) and IXI dataset~\cite{ixi2015dataset}~(PD as reference and T2 as target). All the complex-valued images are cropped to integer multiples of 24 (as the smallest common multiple of the test scale). 
We adopt a commonly used down-sampling treatment to crop the {\itshape k}-space. Concretely, we first converted the original image into the {\itshape k}-space using Fourier transform. Then, only data in the central low-frequency region are kept, and all high-frequency information is cropped out. For the down-sampling factors {\itshape k}, only the central $\frac{1}{k^2}$ frequency information is kept. 
Finally, we used the inverse Fourier transform to convert the down-sampled data into the image domain to produce the LR image.

We compared our Dual-ArbNet with several recent state-of-the-art methods, including two multi-contrast SR methods: McMRSR~\cite{li2022transformer}, WavTrans~\cite{li2022wavtrans}, and three arbitrary scale image SR methods: Meta-SR~\cite{hu2019meta}, LIIF~\cite{chen2021learning}, Diinn~\cite{nguyen2023single}.

\noindent
\textbf{Experimental Setup.} Our proposed Dual-ArbNet is implemented in PyTorch with NVIDIA GeForce RTX 2080 Ti. The Adam optimizer is adopted for model training, and the learning rate is initialized to $10^{-4}$ at the full-training stage for all the layers and decreases by half for every 40 epochs. We randomly extract 6 LR patches with the size of 32$\times$32 as a batch input. Following the setting in~\cite{hu2019meta}, we augment the patches by randomly flipping horizontally or vertically and rotating $90^{\circ}$. The training scale factors of the Dual-ArbNet vary from 1 to 4 with stride 0.1, and the distribution of the scale factors is uniform. The performance of the SR reconstruction is evaluated by PSNR and SSIM.

\begin{table}[tbp]
  \centering
  \setlength\tabcolsep{0.8pt}
  \caption{Quantitative comparison with other methods. Best and second best results are \textbf{highlighted} and \underline{underlined}.}
    \begin{tabular}{|c|c|c|c|c|c|c|c|c|c|}
    \hline
    \multirow{2}{*}{Dataset} & \multirow{2}{*}{Methods} & \multicolumn{4}{c}{In distribution} \vline& \multicolumn{2}{c}{\makecell[c]{Out-of\\distribution}} \vline & \multicolumn{2}{c}{Average} \vline\\
    \cline{3-10}        &       & ×1.5  & ×2    & ×3    & ×4    & ×6    & ×8 & PSNR & SSIM \\
    \hline
    \multirow{6}{*}{fast} & \makecell[l]{McMRSR\cite{li2022transformer}} & 37.773 & \underline{34.546} & 31.087 & 30.141 & 27.859 & 26.200 & 31.268 & 0.889\\
    \cline{2-10}           & \makecell[l]{WavTrans\cite{li2022wavtrans}} & 36.390 & 32.841 & 31.153 & 30.197 & 28.360 & \textbf{26.722} & 30.944 & 0.890\\
    \cline{2-10}           & \makecell[l]{Meta-SR\cite{hu2019meta}}  & 37.243 & 33.867 & 31.047 & 29.604 & 27.552 & 24.536 & 30.642 & 0.880\\
    \cline{2-10}           & \makecell[l]{LIIF\cite{chen2021learning}}     & \underline{37.868} & 34.320 & \underline{31.717} & \underline{30.301} & \underline{28.485} & 26.273 & \underline{31.494} & \underline{0.892}\\
    \cline{2-10}           & \makecell[l]{Diinn\cite{nguyen2023single}}    & 37.405 & 34.182 & 31.666 & 30.243 & 28.382 & 24.804 & 31.114 & 0.887\\
    \cline{2-10}           & \makecell[l]{Ours}     & \textbf{38.139} & \textbf{34.722} & \textbf{32.046} & \textbf{30.707} & \textbf{28.693} & \underline{26.419} & \textbf{31.788} & \textbf{0.896}\\
    \hline
    \multirow{6}{*}{IXI}  & \makecell[l]{McMRSR\cite{li2022transformer}} & 37.450 & 37.046 & 34.416 & 33.910 & 29.765 & 27.239 & 33.304 & 0.914\\
    \cline{2-10}           & \makecell[l]{WavTrans\cite{li2022wavtrans}}& 39.118 & \underline{38.171} & \underline{37.670} & \underline{35.805} & \underline{31.037} & \underline{27.832} & \underline{34.940} & \underline{0.958}\\
    \cline{2-10}           & \makecell[l]{Meta-SR\cite{hu2019meta}} & 42.740 & 36.115 & 32.280 & 29.219 & 25.129 & 23.003 & 31.414 & 0.916\\
    \cline{2-10}           & \makecell[l]{LIIF\cite{chen2021learning}} & 41.724 & 36.818 & 33.001 & 30.366 & 26.502 & 24.194 & 32.101 & 0.934\\
    \cline{2-10}           & \makecell[l]{Diinn\cite{nguyen2023single}} & \underline{43.277} & 37.231 & 33.285 & 30.575 & 26.585 & 24.458 & 32.569 & 0.936\\
    \cline{2-10}           & \makecell[l]{Ours}    & \textbf{43.964} & \textbf{40.768} & \textbf{38.241} & \textbf{36.816} & \textbf{33.186} & \textbf{29.537} & \textbf{37.085} & \textbf{0.979}\\
    \hline
    \end{tabular}%

  \label{tab:comparison}%
\end{table}%

\noindent
\textbf{Quantitative Results.} Table~\ref{tab:comparison} reports the average SSIM and PSNR with respect to different datasets under in-distribution and out-of-distribution large scales. Since the SR scale of McMRSR~\cite{li2022transformer} and WavTrans~\cite{li2022wavtrans} is fixed to 2$\times$ and 4$\times$, we use a 2$\times$ model and down-sample the results when testing 1.5$\times$. We use the 4$\times$ model and up-sample the results to test 6$\times$ and 8$\times$, and down-sample the results to test 3$\times$ results. Here, we provide the results with the reference image at HR resolution. As can be seen, our method yields the best results in all datasets. Notably, for out-of-distribution scales, our method performs even significantly better than existing methods. The results confirm that our framework outperforms the state-of-the-art in terms of performance and generalizability.

\begin{figure}[tb]
\centering
\includegraphics[width=1\textwidth]{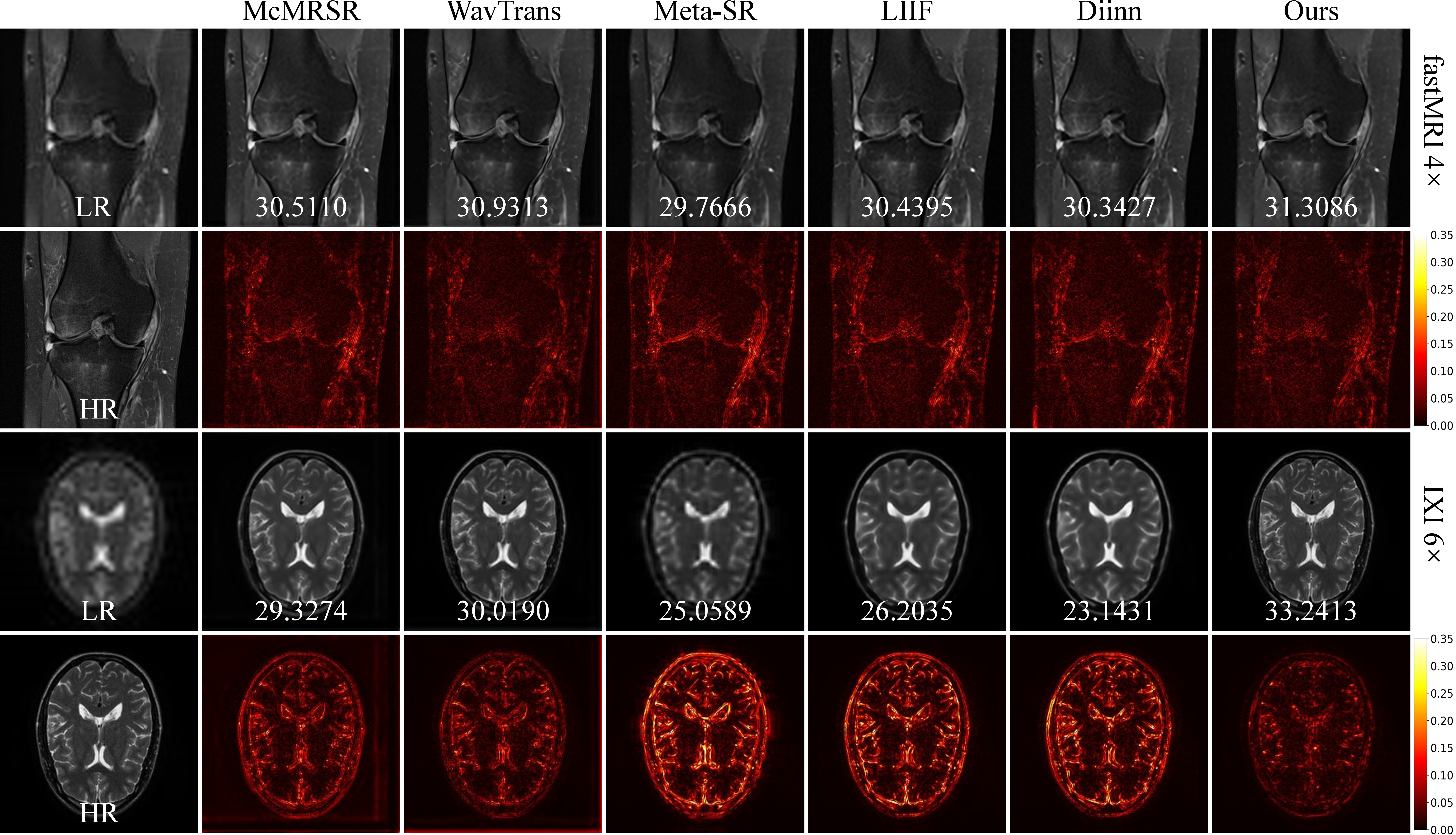}

\caption{Qualitative results and error maps of different SR methods on fastMRI and IXI dataset. The color bar on the right indicates the value of the error map. Our method can reconstruct fewer blocking artifacts and sharper texture details.}
\label{img:results}

\end{figure}

\noindent
\textbf{Qualitative Evaluation.} Figure~\ref{img:results} provides the reconstruction results and the corresponding error maps of the in-distribution scale (4$\times$) and out-of-distribution scale (6$\times$). The more obvious the texture in the error map, the worse the reconstruction means. As can be observed, our reconstructed images can eliminate blurred edges, exhibit fewer blocking artifacts and sharper texture details, especially in out-of-distribution scales.

\begin{table}[htbp]
  \centering
  \setlength\tabcolsep{2.8pt}
  \caption{Ablation study on different training strategies~(top) and key components~(bottom) under fastMRI dataset. Best results are \textbf{highlighted}.}
    \begin{tabular}{|c|c|c|c|c|c|c|c|c|}
    \hline
     \makecell[c]{TrainRef} & \makecell[c]{TestRef} & ×1.5  & ×2    & ×3    & ×4    & ×6    & ×8 &average\\
    \cline{1-9}  LR    & LR    & 37.911 & 34.475 & 31.705 & 30.219 & 28.137 & 24.245 & 31.115 \\
    \cline{1-9}  LR    & HR    & 36.954 & 34.232 & 31.615 & 30.031 & 27.927 & 24.455 & 30.869\\
    \cline{1-9}  HR    & LR    & 35.620 & 33.007 & 30.268 & 28.789 & 26.624 & 24.942 & 29.875\\
    \cline{1-9}  HR    & HR    & 36.666 & 34.274 & 31.916 & \textbf{30.766} & 28.392 & 26.359 & 31.395\\
    \cline{1-9}  Random & LR   & \textbf{38.143} & 34.423 & 31.669 & 30.173 & 27.975 & 25.182 & 31.261\\
    \cline{1-9}  Random & HR   & 38.140 & 34.640 & 32.025 & 30.712 & 28.647 & 26.355 & 31.753\\
    \cline{1-9}  Cur-Random & LR & 38.063 & 34.489 & 31.684 & 30.177 & 28.038 & 25.264 &31.286\\
    \cline{1-9}  Cur-Random & HR & 38.139 & \textbf{34.722} & \textbf{32.046} & 30.707 & \textbf{28.693} & \textbf{26.419} &\textbf{31.788}\\
    \hline
    \end{tabular}%
    \setlength\tabcolsep{0.8pt}
    \begin{tabular}{|c|c|c|c|c|c|c|c|c|c|c|}
      \hline
      Setting     & Ref & Scales & Coord & ×1.5  & ×2    & ×3    & ×4    & ×6    & ×8 & average \\
      \hline
      w/o ref & \ding{55}    & \ding{55}     & \ding{51}     & 37.967 & 34.477 & 31.697 & 30.214 & 28.154 & 24.996 & 31.251\\
      \hline
      w/o scale & \ding{51}     & \ding{55}     & \ding{51}     & 37.951 & 34.663 & 32.063 & 30.681 & 28.623 & 26.413 & 31.732\\
      \hline
      w/o coord & \ding{51}     & \ding{51}     & \ding{55}     & 38.039 & 34.706 & 32.036 & 30.702 & 28.592 & 26.288 &31.727\\
      \hline
      Dual-ArbNet & \ding{51}     & \ding{51}     & \ding{51}     & \textbf{38.139} & \textbf{34.722} & \textbf{32.046} & \textbf{30.707} & \textbf{28.693} & \textbf{26.419} & \textbf{31.788}\\
      \hline
      \end{tabular}%
  \label{tab:random}%
\end{table}%

\noindent
\textbf{Ablation Study on different training strategies.} 
We conduct experiments on different training strategies and reference types to demonstrate the performance of Dual-ArbNet and the gain of Cur-Random, as shown in Table~\ref{tab:random}(top). Regarding the type of reference image, we use HR, LR, Random, Cur-Random for training, and HR, LR for testing. As can be seen, the domain gap appears in inconsistent training-testing pairs, while Random training can narrow this gap and enhance the performance. In addition, the HR-Ref performs better than the LR-Ref due to its rich detail and sharp edges, especially in large scale factors. Based on the Random training, the Cur-Random strategy can further improve the performance and achieve balanced SR results.

\noindent
\textbf{Ablation Study on key components.} In Table~\ref{tab:random}(bottom), to evaluate the validity of the key components of Dual-ArbNet, we conducted experiments without introducing coordinate information, thus verifying the contribution of coordinate in the IDF, named w/o coord. The setting without introducing scale factors in implicit decoding is designed to verify the effect of scale factors on model performance, named w/o scale. To verify whether the reference image can effectively provide auxiliary information for image reconstruction and better restore SR images, we further designed a single-contrast variant model without considering the reference image features in the model, named w/o ref. All the settings use Cur-Random training strategy.

As can be seen that the reconstruction results of w/o coord and w/o scale are not optimal because coordinates and scale can provide additional information for the implicit decoder. We observe that w/o ref has the worst results, indicating that the reference image can provide auxiliary information for super-resolving the target image.

\section{Conclusion}
In this paper, we proposed the Dual-ArbNet for MRI SR using implicit neural representations, which provided a new paradigm for multi-contrast MRI SR tasks. It can perform arbitrary scale SR on LR images at any resolution of reference images. In addition, we designed a new training strategy with reference to the idea of curriculum learning to further improve the performance of our model. Extensive experiments on multiple datasets show that our Dual-ArbNet achieves state-of-the-art results both within and outside the training distribution. We hope our work can provide a potential guide for further studies of arbitrary scale multi-contrast MRI SR.

\subsubsection{Acknowledgements} 
This work was partly supported by the National Natural Science Foundation of China (Nos. 62171251 \& 62311530100), the Special Foundations for the Development of Strategic Emerging Industries of Shenzhen (Nos.JCYJ20200109143010272 \& CJGJZD20210408092804011) and Oversea Cooperation Foundation of Tsinghua.
\nocite{*}
\bibliography{myReference}
\bibliographystyle{splncs04}

\clearpage 
\begin{appendix}
\section*{Appendix}

\begin{table}[h]
\caption{Details of the two datasets we used. The number indicates the target-reference image pairs we use for training, testing and validation.}\label{tab1}
\centering
\setlength{\tabcolsep}{4mm}
\begin{tabular}{|l|c|c|}
\hline
Datasets               &  fastMRI & IXI  \\ \hline
Reference modality     &  PD      & PD   \\ \hline
Target modality        &  FS-PD   & T2   \\ \hline
Train$\backslash$Valid$\backslash$Test & 488$\backslash$78$\backslash$100 & 890$\backslash$142$\backslash$182\\
\hline
\end{tabular}
\end{table}

\begin{table}[h]
\caption{Setting of each stage in Cur-Random strategy. Arbitrary means scales are random select form 1 to 4 with interval 0.1, LR/HR-Ref means the resolution of the reference image is the same as the target LR/HR image. }\label{tab:cur-learning}
\centering
\setlength{\tabcolsep}{4mm}
\begin{tabular}{|l|c|c|c|}
\hline
Stage           & Target Scale & Ref Scale & learning rate                  \\ \hline
warm-up        &  2x,3x,4x    & HR        & $5\times 10^{-5}$   \\ \hline
pre-learning    &  arbitrary   & HR        & $1\times 10^{-4}$   \\ \hline
full-training   &  arbitrary   & LR,HR     & Step LR             \\ \hline
\end{tabular}
\end{table}


\begin{table}[h]
  \centering
  \setlength\tabcolsep{3pt}
  \caption{Ablation study on w/o k-Loss. Best results are \textbf{highlighted}. Introducing k-Loss can further improve the average performance of our Dual-ArbNet.}
    \begin{tabular}{|c|c|c|c|c|c|c|c|c|c|c|}
    \hline
    Setting     & K-Loss & ×1.5  & ×2    & ×3    & ×4    & ×6    & ×8 & average \\
    \hline
    w/o k-loss & \ding{55} & \textbf{38.187} & 34.679 & 32.011 & 30.662 & 28.631 & 26.374 & 31.757\\
    \hline    
    Dual-ArbNet & \ding{51} & 38.139 & \textbf{34.722} & \textbf{32.046} & \textbf{30.707} & \textbf{28.693} & \textbf{26.419} & \textbf{31.788}\\
    \hline
    \end{tabular}%
  \label{tab:ablation}%
\end{table}

\begin{figure}[h]
\centering
\includegraphics[width=0.9\textwidth]{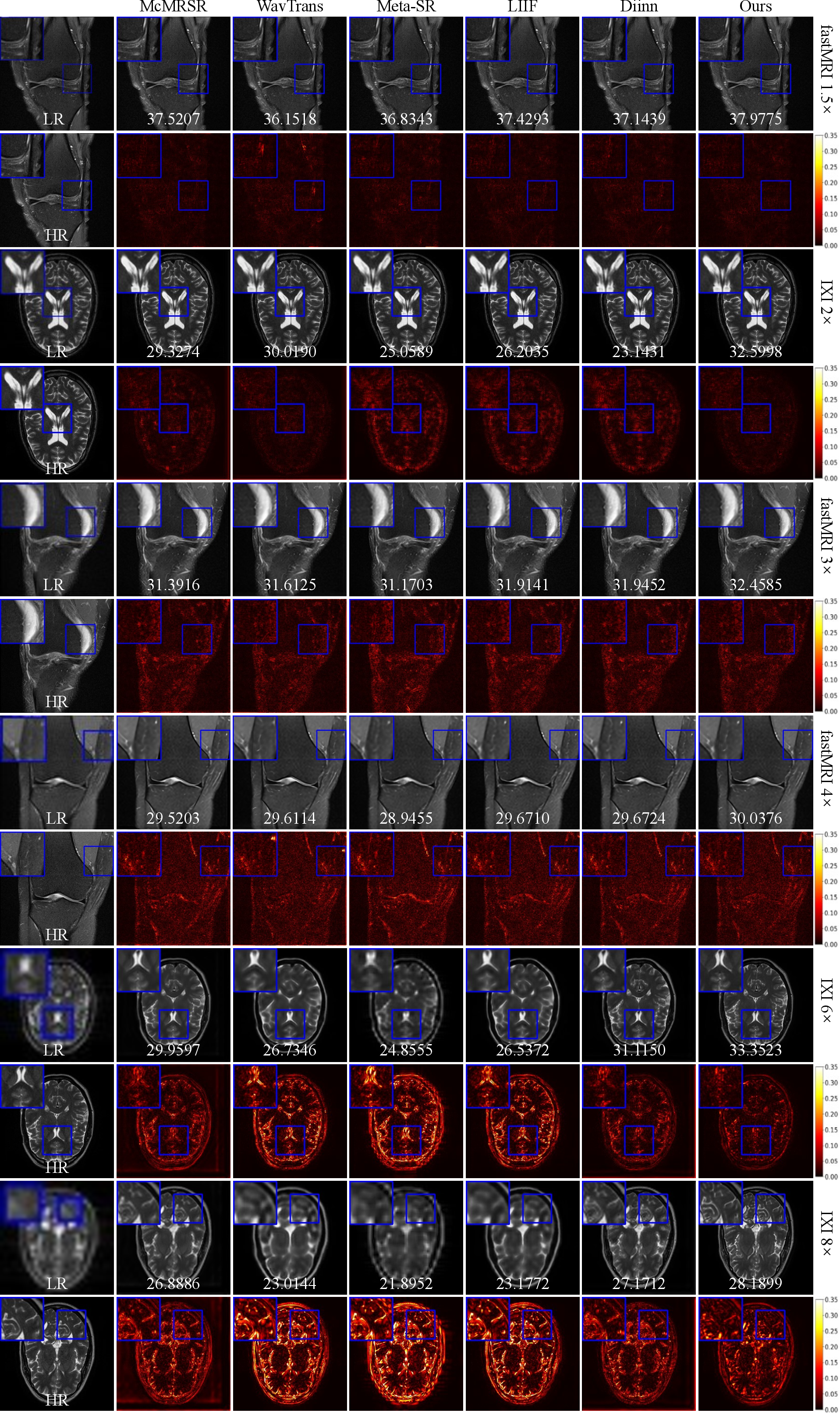}
\caption{Qualitative results (PSNR(dB)) and error maps of different SR methods on fastMRI and IXI dataset. The color bar on the right indicates the value of the error map. Our method can reconstruct fewer blocking artifacts and sharper texture details.}
\label{img:resultin}
\end{figure}

\end{appendix}

\end{document}